# Science Visualization and Discursive Knowledge*

Loet Leydesdorff

University of Amsterdam, Amsterdam School of Communications Research (ASCoR)
Kloveniersburgwal 48, 1012 CX Amsterdam, The Netherlands
loet@leydesdorff.net ; http://www.leydesdorff.net

**Abstract**
Positional and relational perspectives on network data have led to two different research traditions in textual analysis and social network analysis, respectively. Latent Semantic Analysis (LSA) focuses on the latent dimensions in textual data; social network analysis (SNA) on the observable networks. The two coupled topographies of information-processing in the network space and meaning-processing in the vector space operate with different (nonlinear) dynamics. The historical dynamics of information processing in observable networks organizes the system into instantiations; the systems dynamics, however, can be considered as self-organizing in terms of fluxes of communication along the various dimensions that operate with different codes. The development over time adds evolutionary differentiation to the historical integration; a richer structure can process more complexity.

## Introduction

The visualization of scientific developments using different types of software and algorithms is booming business on the Internet. Recently, a comprehensive *Atlas of Science* was also published (Börner, 2010). Unlike the geographic map, however, science has no natural baselines. Scientific domains can be spanned in different directions, such as in terms of (inter)disciplines (Small & Garfield, 1985). In a visualization one tries to capture those complex dynamics by reducing images to two-dimensional maps or three-dimensional landscapes. Furthermore, if the time-axis is involved as in an animation of evolving dynamics, additional constructions are needed for stabilizing the representation so that the results can be captured as a mental map (Liu & Stasko, 2010; Misue *et al*., 1995).

The sciences span intellectual spaces which can be mapped in terms of words (e.g., title-words), authors, or their co-occurrences (Callon *et al*., 1983; White & Griffith, 1982; White & McKain, 1998). At a higher level of aggregation, journal-journal citation relations—available from the *Science Citation Index*—have been used since the mid-1980s for mapping developments of and among disciplines (Doreian & Fararo, 1985; Leydesdorff, 1986); Tijssen *et al*., 1987). Small and his coauthors further developed the mapping of co-citations (e.g., Garfield, 1978; Small & Sweeney, 1985. Börner *et al*. (2003) provides a review of the visualization of scientific knowledge domains (cf. Leydesdorff, 1987; McKain, 1990).

In this chapter, I argue that observable network relations organize the sciences under study into historical instantiations that can be visualized statically. The development of scholarly discourse, however, can be considered as self-organizing in terms of fluxes of communication along the various dimensions that operate with different (e.g., disciplinary) codes. The development over time adds evolutionary differentiation to the historical integration; a richer structure can process more complexity. Latent Semantic Analysis (LSA) focuses on these latent dimensions in textual data; social network analysis (SNA) on the networks of observable relations. However, the two coupled topographies of information-processing in the network space and meaning-processing in the vector space operate with different (nonlinear) dynamics.

## Multidimensional scaling

Historically, computer-aided visualization of multivariate data predated the advent of the personal computer and the Internet. Based on Kruskall (1964), scholars in psychometrics developed spatial representations of sets of variables by multidimensional scaling (MDS; e.g., Kruskall & Wish, 1978; Schiffman *et al*., 1981). Among other forms of output, MDS can generate a two-dimensional map. The first large-scale MDS program ALSCAL ("alternating least square analysis") is still available in current versions of statistical packages such as SPSS.

**Table 1.** Flying mileages among ten American Cities

| | Atlanta | Chicago | Denver | Houston | Los Angeles | Miami | New York | San Francisco | Seattle | Washington DC |
|---|---|---|---|---|---|---|---|---|---|---|
| Atlanta | 0 | . | . | . | . | . | . | . | . | . |
| Chicago | 587 | 0 | . | . | . | . | . | . | . | . |
| Denver | 1212 | 920 | 0 | . | . | . | . | . | . | . |
| Houston | 701 | 940 | 879 | 0 | . | . | . | . | . | . |
| Los Angeles | 1936 | 1745 | 831 | 1374 | 0 | . | . | . | . | . |
| Miami | 604 | 1188 | 1726 | 968 | 2339 | 0 | . | . | . | . |
| New York | 748 | 713 | 1631 | 1420 | 2451 | 1092 | 0 | . | . | . |
| San Francisco | 2139 | 1858 | 949 | 1645 | 347 | 2594 | 2571 | 0 | . | . |
| Seattle | 2182 | 1737 | 1021 | 1891 | 959 | 2734 | 2408 | 678 | 0 | . |
| Washington DC | 543 | 597 | 1494 | 1220 | 2300 | 923 | 205 | 2442 | 2329 | 0 |

Table 1 provides distances in terms of flying mileages among ten American cities (SPSS, 1993; Leydesdorff & Vaughan, 2006). MDS enables us to regenerate the map from which these distances were obtained by minimizing the stress *S* in the projection (Figure 1). Feeding this data into ALSCAL, for example, leads not surprisingly to an almost perfect fit (*S* = 0.003).

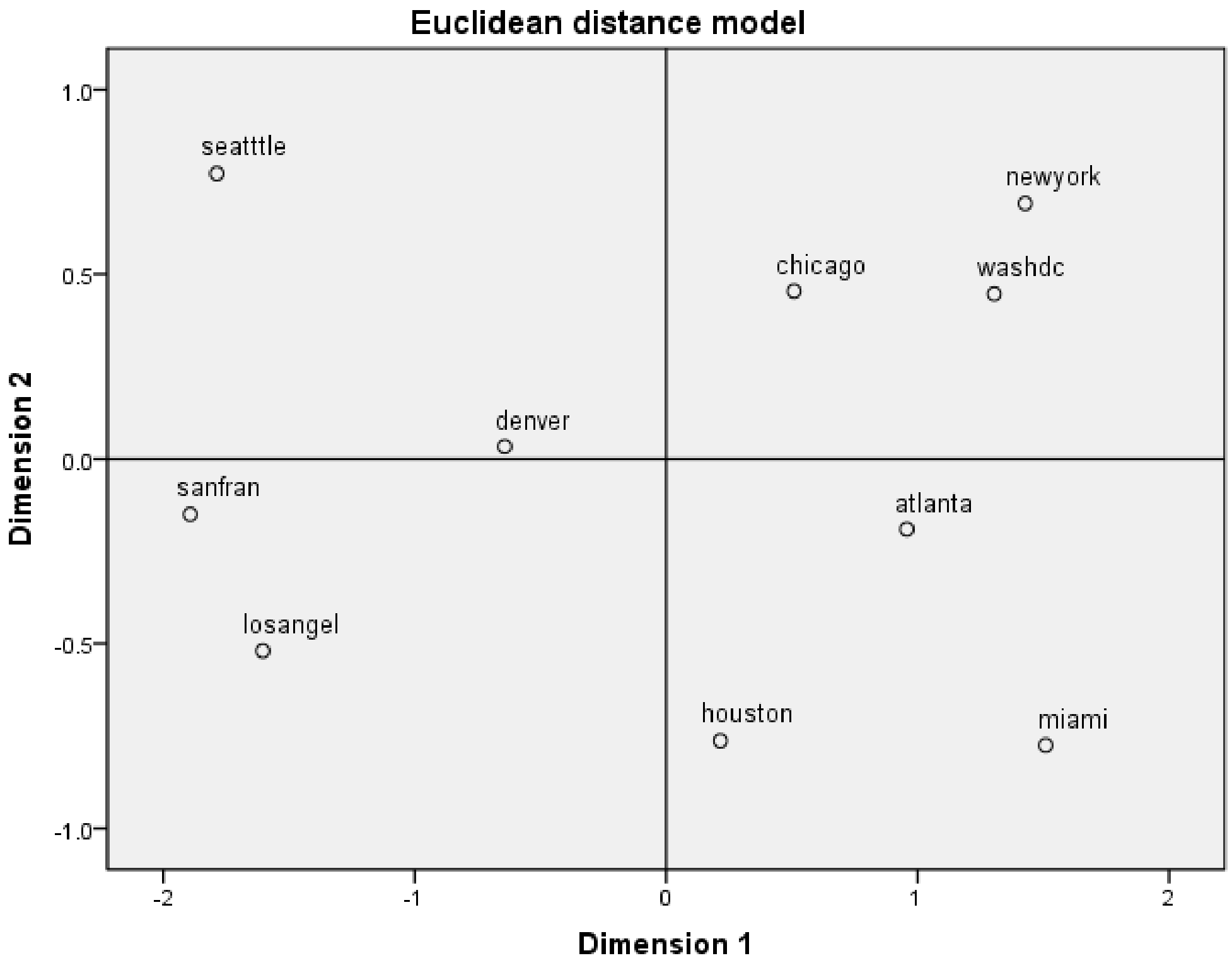

**Figure 1:** MDS mapping (ALSCAL) of ten American cities using the distance matrix in Table 1 (normalized raw stress = 0.003).

This data measures *dissimilarity*, as the larger the numbers, the further apart the cities are, i.e., the more "dissimilar" they are in location. One can also use similarity measures for mapping, such as correlation coefficients. Options that might be added to a next generation of such maps can be listed as follows:

1. In addition to the position of the variable names, one would like to be able to visualize the network of connections among the cities;
2. Other measures of distance than Euclidean ones; for example, correlations in a multidimensional (vector) space provide a different topology;
3. Groupings of nodes using different colors based on attribute values;

4. Nodes and links can be scaled with the values of attributes; etc.

A large number of network visualization and analysis programs nowadays provide these features and can be downloaded from the Internet.

## Graph theory and network analysis

During the 1980s, graph theory became available as a theoretical basis for network analysis. In the original programs (such as GRADAP) the links had to be drawn by hand. UCINet 2.0 (1984) provided the first network analysis program that integrated a version of MDS (MINISSA),[1] but the capacity of the number of variables was at the time limited to 52: 26 upper-case and 26 lower-case characters could be indicated (Freeman, 2004). These programs allowed for using similarity measures other than Euclidean distances. For example, Leydesdorff (1986) used Pearson correlations to visualize factor structures in aggregated journal-journal citation matrices using UCINet 2.0.

Graphic interfaces became available during the 1990s with the further development of Windows (Windows'95) and the Apple computers. Pajek followed as a visualization and analysis tool for large networks in 1996 (De Nooy *et al*., 2005). Pajek also allows for non-Western characters such as Chinese and Arabic (Leydesdorff & Jin, 2005).[2]

---

[1] MINISSA is an acronym for "Michigan-Israel-Nijmegen Integrated Smallest Space Analysis"; it became available around 1980 (Schiffman *et al*., 1981).

[2] Pajek is a freeware program for network visualization and analysis available at http://vlado.fmf.uni-lj.si/pub/networks/pajek/.

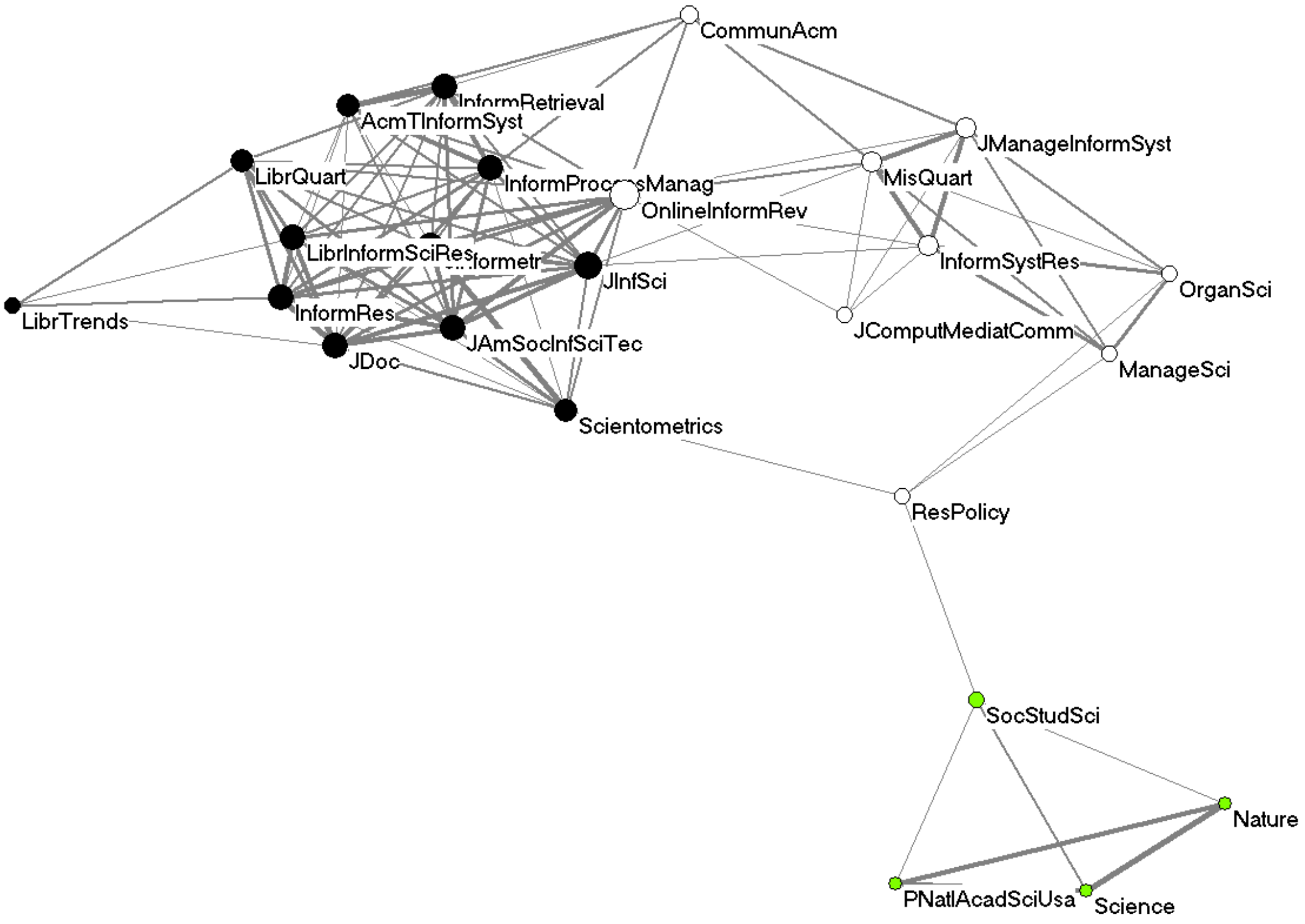

**Figure 2**: 25 journals most cited by authors in *JASIST* during 2010; Kamada & Kawai (1989) used for the layout; node sizes proportionate to degree centrality; node colors according to modularity ( $Q$ = 0.328); edge width proportionate to *cosine* values (*cosine* > 0.2).

Figure 2 provides an example of the current state of the art: the aggregated citation network of the *Journal of the American Society for Information Science and Technology* (*JASIST*) as mapped in 2010. (These 25 journals are cited in *JASIST* to the extent of more than 1% of its total citations.) The matrix is analyzed using both Pajek and Gephi;[3] links are indicators of cosine-similarities between the citing patterns of these journals; the vertices are colored according to the modularity algorithm ($Q$ = 0.328; Blondel *et al*., 2008), and sized according to their degree centrality (De Nooy *et al*., 2005).

[3] Gephi is an open-source program for network analysis and visualization, available at https://gephi.org/.

*Research Policy,* positioned between the three components in this map, has accordingly the highest betweenness centrality (0.305). Although different in some details, both the factor analysis[4] and the modular decomposition classify *Research Policy* as belonging to the information-systems group of journals (within this context!). The visualization adds a network of relations among the nodes. As noted, one is able to use attributes of nodes and links in order to further enrich the visual.

**Relational and positional maps of science**

Using MDS one visualizes the variables as a system (e.g., a word-document matrix). In spatial terms, the words attributed to documents are considered as vectors that are vector-summed into a vector space (Salton & McGill, 1983). Given parameter choices (such as the similarity measure), the projection of the variables in MDS is deterministic. The Euclidean distance between San Francisco and New York, for example, does not change depending on the intensity of the network relations (e.g., flights) between these two cities.

In network analysis, one is often more interested in a representation that uses the intensity of the relations as the distance on the map. For example, two authors who often coauthor should be positioned next to each other in a co-authorship map. In this case, it is not the *correlations* among the distributions, but the *relations* among the nodes that are used for the mapping. Graph-analytic algorithms (e.g., Kamada & Kawai, 1989) optimize the network in terms of relations.

[4] Three factors explain 49.2% of the variance in this matrix.

The choice of starting-point can be random, and each run may hence lead to a somewhat different outcome.

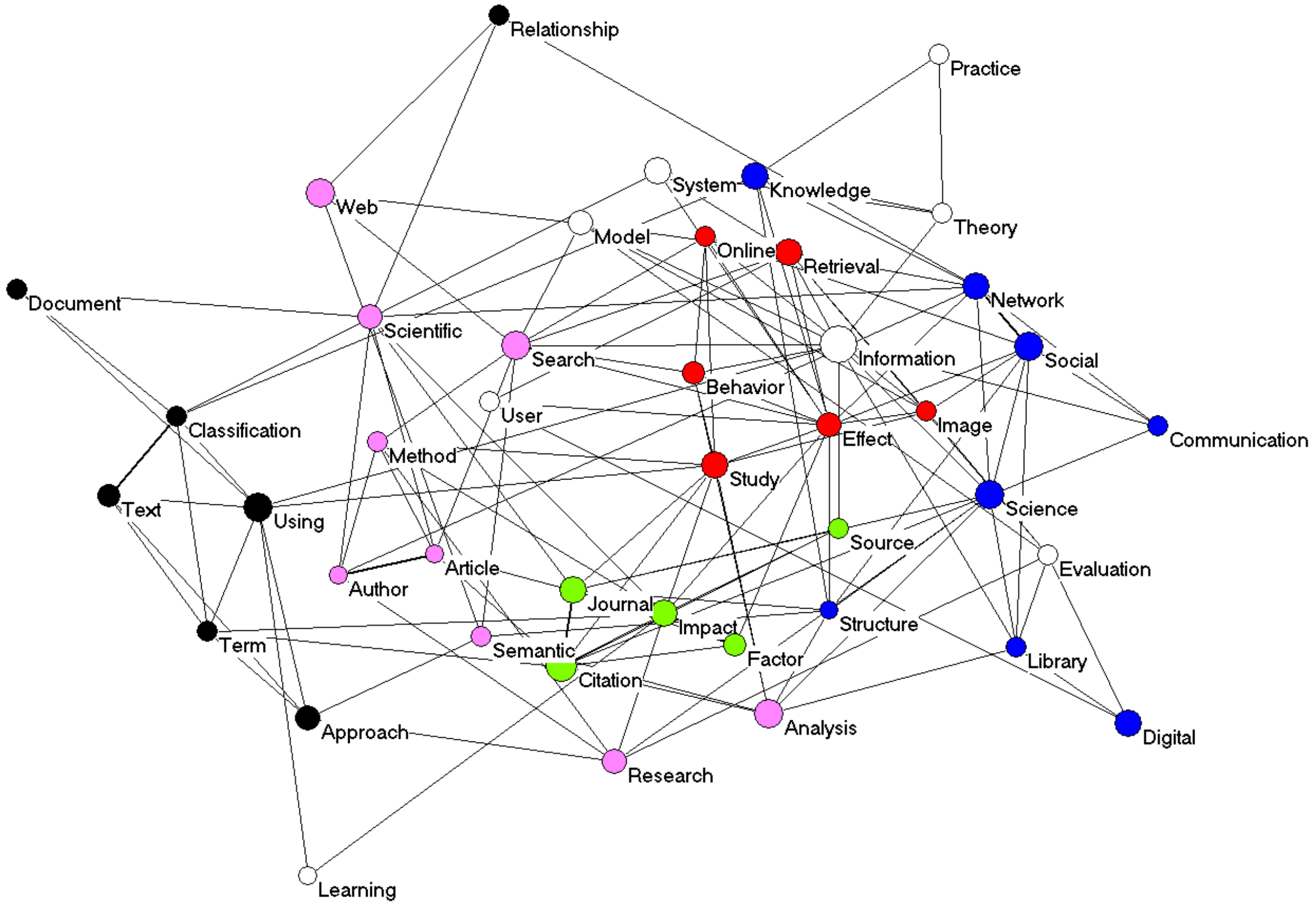

**Figure 3**: Cosine-normalized map of 43 words occurring more than 10 times during 2010 and 2011 in titles of *JASIST*. (*Cosine* ≥ 0.1; Kamada & Kawai, 1989.) The nodes are colored according to the five-factor solution of this network (Varimax rotated; SPSS), and scaled in accordance to their degree centrality.

Let us compare the two approaches of optimizing the vector space versus the network topology. In Figures 3 and 4, 43 title words are included that occurred more than ten times among the 455 titles in the 2010 and 2011 volumes of the *Journal of the American Society for Information Science and Technology* (*JASIST*). A five-factor solution in the underlying data matrix is used for coloring the nodes in the vector space (Figure 3) and the network space (Figure 4), respectively.

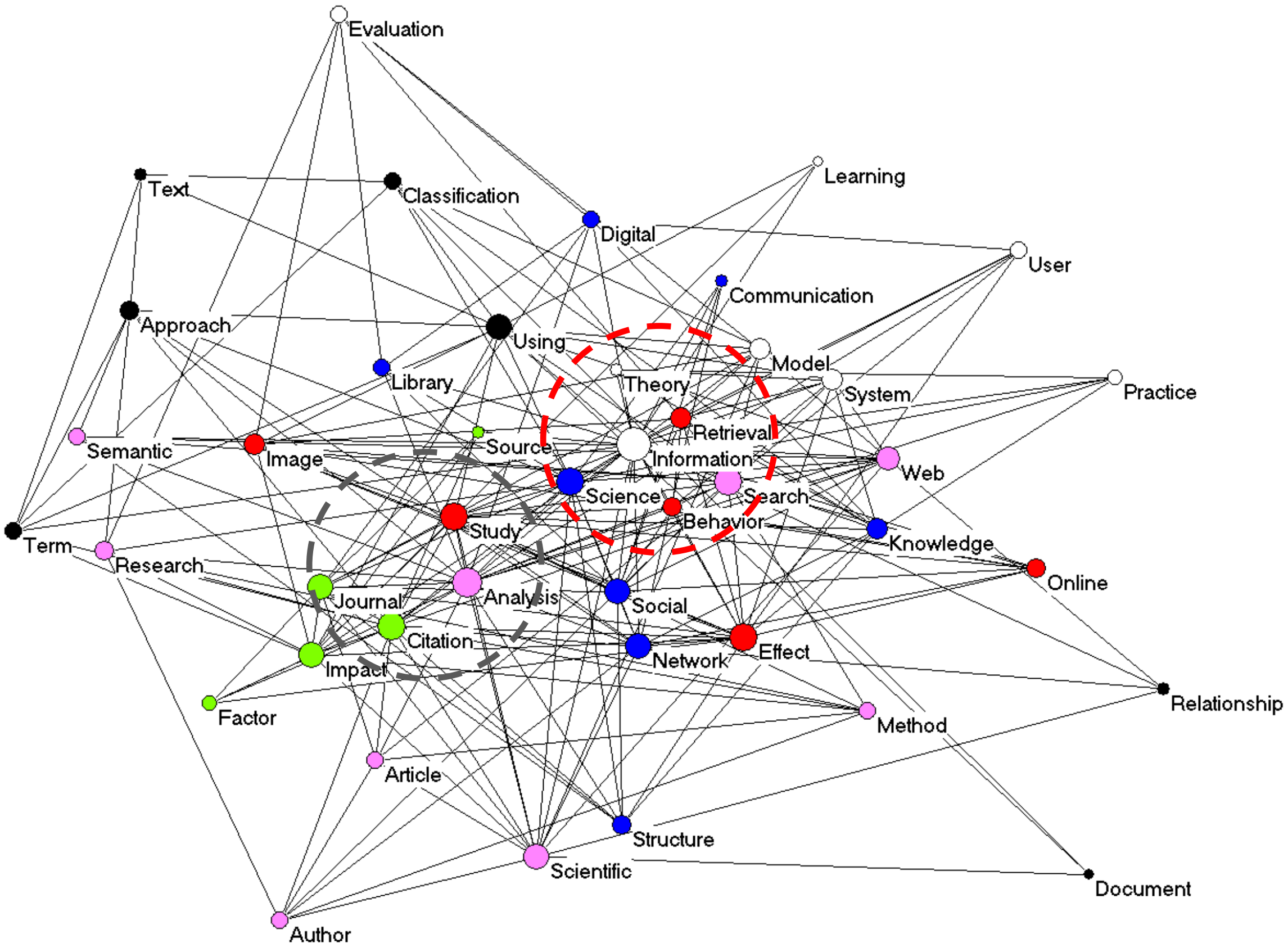

**Figure 4**: Co-occurrence map of 43 words occurring more than 10 times during 2010 and 2011 in titles of *JASIST*. (Co-ocurrence values ≥ 2; Kamada & Kawai, 1989; nodes scaled with degree centrality.)

Factor 1, for example, is composed of the words “impact,” “factor,” “journal,” “citation,” and “source.” These (green-colored) words are grouped in both figures: they not only entertain strong relations to one another (Figure 4), but also co-occur in similar *patterns* among the other title-words in the sample (Figure 3). Factor 4, however, with primary factor loadings for the words “effect,” “image,” “study,” “online,” and “behavior,” can more easily be distinguished in Figure 3 than Figure 4. These words co-occur with other words in the set more diffusely, yet they form a latent dimension of the data.

In other words, there is no necessary relationship between co-occurrences in the observable network of relations, and correlations among co-occurrence patterns. The co-occurrence patterns can be mapped using the correlation coefficients among the distributions, whereas the values of co-occurrence relations provide us with a symmetrical (affiliations) matrix that can be visualized directly. In the latter case one visualizes the network of observable *relations,* whereas in the former one visualizes the latent structure in this data. Two synonyms, for example, may have (statistically) similar *positions* in a semantic map, but they will rarely co-occur in a single title.

These two perspectives on the data have led to two different research traditions in textual analysis and social network analysis, respectively. Latent Semantic Analysis (LSA) focuses on the latent dimensions in textual data; social network analysis (SNA) on the observable relations in networks. In SNA, for example, eigenvector-centrality—that is, factor loading on the first factor—can be used as an attribute of the nodes, whereas in LSA the factors (eigenvectors) in different directions organize the semantic maps (Landauer *et al*., 1998). The factor-analytic approach has been further developed using Singular Value Decomposition (SVD), whereas graph theory has provided an alternative paradigm for developing algorithms in SNA.

A star in a graph can be in the center of the multidimensional space, and therefore not load strongly on any of the dimensions. In Figure 4, for example, the word “information” that occurs 94 times in this set (followed by “citation” occurring only 44 times), did not load positively on any of the five factors distinguished; this variable is factor-neutral and therefore colored white. However, using the degree distribution for sizing the nodes in Figure 4, “information” has the highest degree, co-occurring with 37 of the 43 title-words, followed by “analysis” with a degree of 33. A core set of words surrounding "information" (circled red in Figure 4) belongs to the center of the field of the information sciences. "Citation" (Factor 1) and "Analysis" (Factor 3) are part of a secondary grouping of the relations (grey circled).

### Interpreting science visualizations

When a network is spanned in terms of relations, this process shapes an architecture in which all components have a position. The analysis of this architecture (that is, the set of relations) enables us to specify what the relations mean in the network as a system. For example, the word "information" was most central in this network (Figure 4), but it was not colored in terms of having meaning in any of the relevant dimensions indicated at the systems level. Yet, the word as a variable carries Shannon-type information (uncertainty; Shannon, 1948).

The graph-analytical approach informs us as analysts about the network of relations, but not about what these relations mean in terms of the discourse(s) under study. However, graph-theoretical concepts such as centrality also have meaning in social network analysis. The analyst's (meta-)discourse can be distinguished from the communication among the words under study. The latter communications can represent scholarly discourses, political discourses, and/or newspaper information.

Within each of these discourses, codes of communication can span dimensions that provide the communicated words with meaning. Both the developments in the observable networks (vectors) and the hypothesized dimensions (eigenvectors) can be theorized. The relations among nodes can be considered as attributes of the nodes, but the dimensions of the communication are attributes of the links. SNA focuses on the positions of nodes in terms of vectors, whereas LSA focuses on the position of links in terms of these next-order structures. This scheme can be generalized: the relations among authors can also be considered as a system of links and therefore another semantic domain. Any system that can position its components as a system, provides itself and its elements thereby with meaning (Maturana, 1978). A discourse, for example, provides meaning to the words that are communicated.

The two perspectives of meaning-processing and information-processing can be considered as feedback mechanisms operating upon each other . The shaping of the networks of relations causes structures that can feed back evolutionarily as a next-order system upon the networks of relations from which they emerge. Meaning is provided from the perspective of hindsight, but with reference to other possibilities ("horizons of meaning;" cf. Husserl, 1929). The next-order meaning-processing cannot continue without information-processing; otherwise, the systems would no longer be historical. The historical instantiation can from this perspective be considered as a retention mechanism of the semantic systems that evolve over time (Leydesdorff, 2011a).

**The network and the vector space**

The multidimensional (vector) space can be considered as a system of relations including interaction terms; the network space as an aggregate of observable relations among nodes. One can also call the network relations first-order (since observable) and the vector space second-order because the latent dimensions of the system are not given, but hypothesized; for example, in a factor-analytical model. Whereas observable variation is stochastic, latent structure is deterministic. The deterministic selection mechanism(s), however, can be expected to be further developed over time in parallel to the networks of relations because of the feedback mechanisms involved.

Accordingly, the systems view of MDS is deterministic, whereas the graph-analytic approach can also begin with a random or arbitrary choice of a starting point. Using MDS, the network is first conceptualized as a multi-dimensional space that is then reduced stepwise to lower dimensionality. At each step, the stress increases; Kruskall's stress function is formulated as follows:

$$S = \sqrt{\frac{\sum_{i \neq j} (\|x_i - x_j\| - d_{ij})^2}{\sum_{i \neq j} d_{ij}^{\ 2}}} \qquad (1)$$

In this formula $\|x_i - x_j\|$ is equal to the distance on the map, while the distance measure $d_{ij}$ can be, for example, the Euclidean distance in the data under study. As noted, one can use MDS to illustrate factor-analytic results in tables, and in this case the Pearson correlation obviously provides the best match.

Spring-embedded or force-based algorithms can be considered as a generalization of MDS, but were inspired by the above-mentioned developments in graph theory during the 1980s. Kamada and Kawai (1989) were the first to reformulate the problem of achieving target distances in a network in terms of energy optimization. They formulated the ensuing stress in the graphical representation as follows:

$$S = \sum_{i \neq j} s_{ij} \quad \text{with} \quad s_{ij} = \frac{1}{d_{ij}^{\,2}} (\|x_i - x_j\| - d_{ij})^2 \qquad (2)$$

Equation 2 differs from Equation 1 by taking the square root in Equation 1, and because of the weighing of *each* term with $1/d_{ij}^2$ in the numerator of Equation 2. This weight is crucial for the quality of the layout, but defies normalization with $\sum d_{ij}^2$ in the denominator of Equation 1; hence the incomparability between the two stress values.

The ensuing difference at the conceptual level is that spring-embedding is a graph-theoretical concept developed for the topology of a network. The weighing is achieved for each individual link. MDS operates on the multivariate space as a system, and hence refers to a different topology. In the multivariate space, two points can be close to each other without entertaining a relationship (Granovetter, 1973). For example, they can be close or distanced in terms of the correlation between their *patterns* of relationships (cf. Burt, 1992).

In the network topology, Euclidean distances and geodesics (shortest distances) are conceptually more meaningful than correlation-based measures. In the vector space, correlation analysis (factor analysis, etc.) is appropriate for analyzing the main dimensions of a system. The cosines of the angles among the vectors, for example, build on the notion of a multi-dimensional space. In bibliometrics, Ahlgren *et al*. (2003) have argued convincingly in favor of the cosine as a non-parametric similarity measure because of the skewedness of the citation distributions and the abundant zeros in citation matrices. Technically, one can also input a cosine-normalized matrix into a spring-embedded algorithm. The value of (1 – *cosine*) is then considered as a distance in the vector space (Leydesdorff & Rafols, 2011). In sum, there is a wealth of possible combinations in a parameter space of clustering algorithms and similarity criteria.

**The visualization of heterogeneous networks**

The two coupled topographies of information-processing in the network space and meaning-processing in the vector space operate with different (nonlinear) systems dynamics (Luhmann, 1995). The historical dynamics of information processing in instantiations organizes the system, and thus interfaces and tends to integrate the (analytically orthogonal) dynamics along each eigenvector. The systems dynamics, however, can be considered as self-organizing in terms of fluxes along the various dimensions—used as codifiers of the communication—and with potentially different speeds. This development over time adds evolutionary differentiation to the historical integration; a richer structure can process more complexity.

Integrating retention can be organized in dimensions other than differentiating expansion. For example, archives and reflexive authors historicize and thus stabilize the volatile networks of new ideas, metaphors, and concepts. Relations among words can be considered as providing us with access to the variation, whereas cited references anchor new knowledge claims in older layers of texts (Lucio-Arias & Leydesdorff, 2009). Authors and institutions may provide historical stability because differences are reflected and locally integrated in communicative actions.

The textual domain provides us with options to combine these different layers in visualizations and animations. The sciences evolve as heterogeneous networks of words, references, authors, and at different levels of aggregation. The composing subdynamics, for example, of specialties and disciplines are not organized neatly in terms of specific variables, but in terms of configurations of variables, such as specific resonances among cognitive horizons (paradigms), social identities, and corpora of literature. The human beings involved (and their organizations) cannot be reduced to literature, and cognitive development can be considered as a latent dimension emerging in networks of texts and people (Leydesdorff, 1998).This thesis of the heterogeneity of the techno-sciences was first proposed by authors in the semiotic tradition (Callon *et al*., 1983).

Because the different dynamics at interfaces within and between knowledge-based systems (such as science, technology, and innovation) are documented in texts, the texts can provide us with access to the different dimensions. In SNA, for example, these various dimensions of the data can be mapped as modalities. Another option for mapping hybrid networks was suggested by Leydesdorff (2010). All relevant variables can be attributed to (sets of) documents as units of analysis. The various asymmetrical matrices of *n* documents versus, for example, *k* words and *m* authors can be aggregated as visualized in Figure 5.

| | $au_1$ | $au_2$ | … | $au_m$ |
|---|---|---|---|---|
| $doc_1$ | $a_{11}$ | $a_{21}$ | … | $a_{m1}$ |
| $doc_2$ | $a_{12}$ | … | … | $a_{m2}$ |
| $doc_3$ | … | … | … | … |
| … | … | … | … | … |
| … | … | … | … | … |
| $doc_n$ | $a_{1n}$ | … | … | $a_{mn}$ |

+

| | $w_1$ | $w_2$ | … | $w_k$ |
|---|---|---|---|---|
| $doc_1$ | $b_{11}$ | $b_{21}$ | … | $b_{k1}$ |
| $doc_2$ | $b_{12}$ | … | … | $b_{k2}$ |
| $doc_3$ | … | … | … | … |
| … | … | … | … | … |
| … | … | … | … | … |
| $doc_n$ | $b_{1n}$ | … | … | $b_{kn}$ |

=

| | $v_1$ | $v_2$ | … | $v_{(m+k)}$ |
|---|---|---|---|---|
| $doc_1$ | $c_{11}$ | $c_{21}$ | … | $c_{(m+k)1}$ |
| $doc_2$ | $c_{12}$ | … | … | $c_{(m+k)2}$ |
| $doc_3$ | … | … | … | … |
| … | … | … | … | … |
| … | … | … | … | … |
| $doc_n$ | $c_{1n}$ | … | … | $c_{(m+k)n}$ |

**Figure 5.** Two matrices for *n* documents with *m* authors and *k* words can be combined to a third matrix of *n* documents *versus* (*m* + *k*) variables.

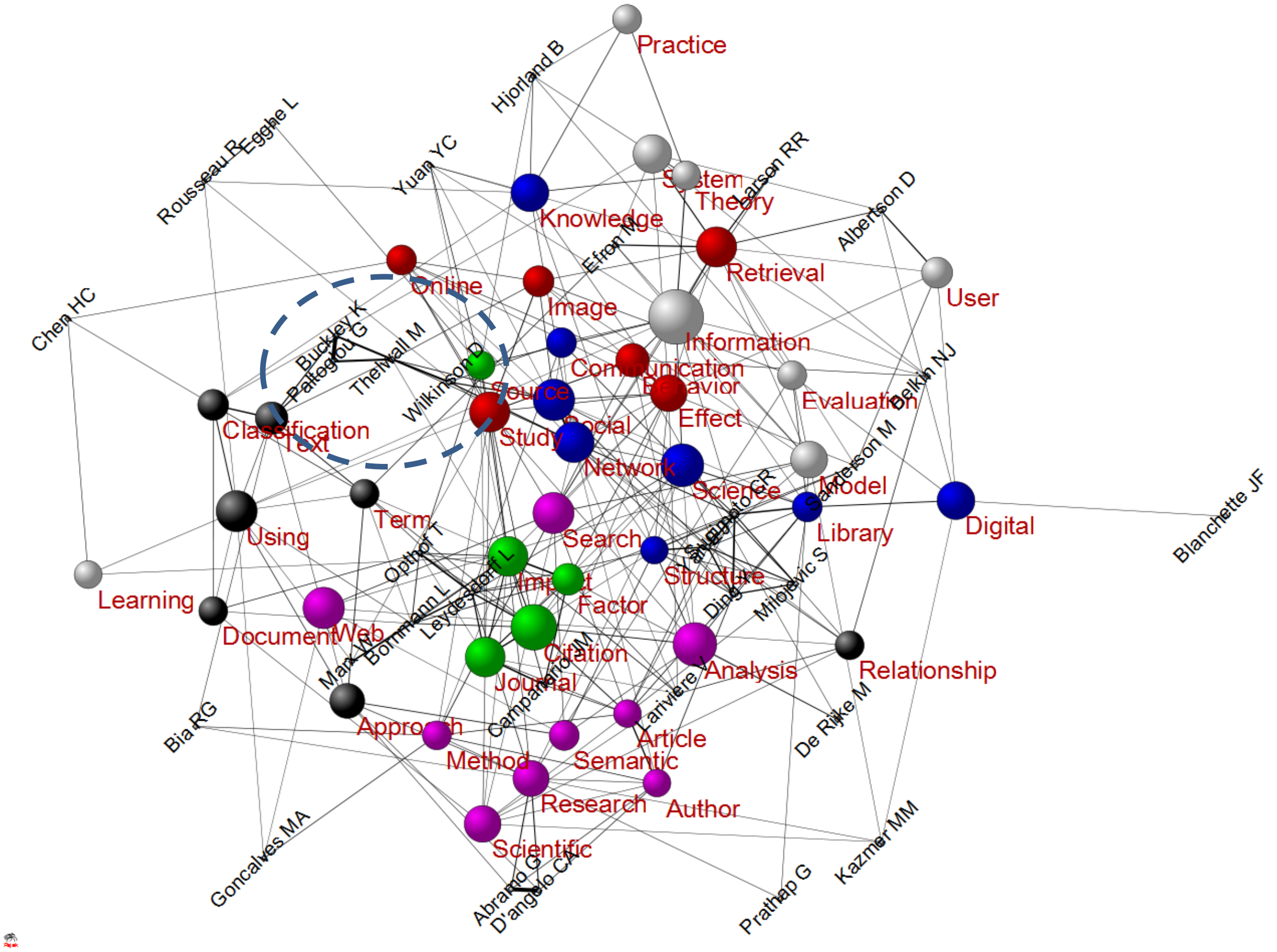

**Figure 6**: 43 words (from Figure 3) and 33 authors related at *cosine* > 0.1.

The resulting matrix can be factor-analyzed or—using matrix algebra—transformed into a symmetrical affiliations matrix. In Figure 6, 33 of the 36 co-authors of these same documents are positioned in a sematic map (as in Figure 3 above). (Three authors were not connected at *cosine* > 0.1). I added a dashed circle around the co-authorship network of Mike Thelwall as an example. Other variables (e.g., cited references, institutional addresses, country names) can be made equally visible, and colored or sized accordingly.

**Animation of the visualizations**

Can the maps for different years (or other time intervals) also be animated? A number of network visualization programs are available that enable the user to smoothen the transitions based on interpolations among the solutions at different moments of time. The dynamic problem is then reduced to a comparatively static one: the differences among maps for different years are assumed to provide us with a representation of the evolution of the system. However, the solution for each year is already an optimization of a higher-dimensional configuration into the two-dimensional plane. It can thus be difficult to distinguish between the development of the system and error.

An analytical solution of the system of partial differential equations provided by all the changing vectors and eigenvectors is impossible, and a numerical one computationally too intensive. Using MDS, however, Gansner *et al*. (2005) proposed minimizing not the stress, but the *majorant* of the stress, as a computationally more effective and methodologically more promising optimization. Baur & Schank (2008) extended this algorithm to layout dynamic networks (cf. Ernten *et al.*, 2004). The corresponding dynamic stress function is provided by the following equation:

$$S = \left[ \sum_t \sum_{i \neq j} \frac{1}{{d_{ij,t}}^2} \left( \left\| x_{i,t} - x_{j,t} \right\| - d_{ij,t} \right)^2 \right] + \left[ \sum_{1 \le t < |T|} \sum_i \omega \left\| x_{i,t} - x_{i,t+1} \right\|^2 \right] \qquad (3)$$

In Equation 3, the left-hand term is equal to the static stress, while the right-hand term adds the dynamic component, namely the stress over subsequent years. This dynamic extension penalizes drastic movements of the position of node *i* at time *t* ($\vec{x}_{i,t}$) toward its next position ($\vec{x}_{i,t+1}$) by increasing the stress value. Thus, stability is provided in order to preserve the mental map between consecutive layouts (Liu & Stasko, 2010).

In other words, the configuration for each year can be optimized in terms of the stress in relation to the solutions for previous years and in anticipation of the solutions for following years. In principle, the algorithm allows us (and the dynamic version of *Visone*—available at http://www.leydesdorff/visone—enables us) to extend this method to more than a single time step. Using a single year in both directions, Leydesdorff & Schank (2008) animated, for example, the aggregated journal-journal citations in "nanotechnology" during the transition of this field at the end of the 1990s (available at http://www.leydesdorff.net/journals/nano).

Note that this approach is different from taking the solution for the previous moment in time as a starting position for a relative optimization. The nodes are not repositioned given a previous configuration, but the previous and the next configurations are included in the algorithmic analysis for each year. More recently, Leydesdorff (2011b) further elaborated this approach by projecting the eigenvectors as constructs among the variables into the animations (e.g., at http://www.leydesdorff.net/eigenvectors/commstudies/). Thus, one can make visible not only the evolution of observable variables, but also the evolution of latent structures. In principle, it would be possible to decompose the resulting stress into dynamic and static components.

## Conclusion and future directions

The relations between semantic maps and social networks have been central to my argument because when visualizing the sciences as bodies of knowledge, the multi-modal network of words, authors, etc., has to be specified. Discursive knowledge is communicated, and thus a network visualization is possible in different dimensions. However, knowledge can be considered as a latent dimension of meaning processing in a network: discursive knowledge emerges in configurations of words, authors, references, etc., and can then be codified and institutionalized, for example, in journals, specialties, departments, and disciplines. The self-organization of the sciences in latent dimensions conditions and enables the observable relations in networks of authors, words, and citation relations.

The sciences are first shaped by the communicating agents, but textual communications can then develop a dynamic of their own as the communications are further codified by theorizing. The sciences develop as systems of rationalized expectations in this codified dimension. However, the developments of ideas leave footprints in the texts (Fujigaki, 1998). The dynamics of texts and authors are different, and the dynamics of communication are (co)determined by the feedback from emerging knowledge dimensions. In Figure 2, for example, the knowledge dimension was operationalized as three groups of journals belonging to different specialties.

The visualization of the sciences as a research program thus requires distinguishing among semantic maps, social networks, and the latent socio-cognitive structures that can emerge on the basis of the interactions among people and texts. Three layers (people, texts, cognitions) coevolve in terms of observable variables and latent eigenvectors. Because of the next-order organization, the variables can be expected to interact among themselves, and to shape and reproduce structures that can both recur on previous states and anticipate further developments of the system(s) (Luhmann, 1995; Maturana, 1978).

Visualization and animation of the sciences are an active research front in the development of the information sciences and bibliometrics. In the future, animations using multiple perspectives can be expected to replace models of multi-variate analysis in which independent factors explain the data. Configurations of variables generate different synergies (Leydesdorff *et al*., in press). These implications follow from considering not only the communication of information, but also its meaning (Krippendorff, 2009; Leydesdorff, 2010); horizons of meaning can be expected to generate redundancy, that is, new and more possibilities that change the value of existing ones.[5]

Animations enable us to capture different perspectives analogously as visualizations capture different arrangements of numbers of variables larger than can be tracked analytically or by using statistics. The development of animations in the coupled layers of information and meaning processing can be expected to raise new questions for the further development of bibliometrics, network analysis, statistics, and relevant neighboring specialisms.

[5] The mutual information in three dimensions ($\mu$; cf. Yeung, 2008, pp. 59f.) among the three main factors structuring the co-word network (Figure 3) is -122.2 mbits, whereas this redundancy virtually disappears when the 33 coauthors are added to the network: $\mu$ = -7.0 mbit (Figure 6). For the social network among the 36 coauthors, this value of $\mu$ is positive. In other words, the coauthor network itself does not communicate meaning in this case (Leydesdorff, 2010, 2011b.).

**Acknowledgements**
I am grateful to Katy Börner for comments on a previous version, and Thomson-Reuters for access to relevant data.